# TOWARDS AN AUTOMATED CLASSIFICATION OF TRANSIENT EVENTS IN SYNOPTIC SKY SURVEYS

S. G. DJORGOVSKI[1,5], C. DONALEK[1], A.A.MAHABAL[1], B. MOGHADDAM[2], M. TURMON[2],
M.J. GRAHAM[3], A.J. DRAKE[3], N. SHARMA[4], Y. CHEN[4]

ABSTRACT. We describe the development of a system for an automated, iterative, real-time classification of transient events discovered in synoptic sky surveys. The system under development incorporates a number of Machine Learning techniques, mostly using Bayesian approaches, due to the sparse nature, heterogeneity, and variable incompleteness of the available data. The classifications are improved iteratively as the new measurements are obtained. One novel feature is the development of an automated follow-up recommendation engine, that suggest those measurements that would be the most advantageous in terms of resolving classification ambiguities and/or characterization of the astrophysically most interesting objects, given a set of available follow-up assets and their cost functions. This illustrates the symbiotic relationship of astronomy and applied computer science through the emerging discipline of AstroInformatics.

## 1. INTRODUCTION

A new generation of scientific measurement systems (instruments or sensor networks) is now generating exponentially growing *data streams*, now moving into the Petascale regime, that can enable significant new discoveries. Often, these consist of phenomena where a rapid change occurs, that have to be identified, characterized, and possibly followed by new measurements in the real time. The requirement to perform the analysis rapidly and objectively, coupled with huge data rates, implies *a need for automated classification and decision making*.

This entails some special challenges beyond traditional automated classification approaches, which are usually done in some feature vector space, with an abundance of self-contained data derived from homogeneous measurements. Here, the input information is generally sparse and heterogeneous: there are only a few initial measurements, and the types differ from case to case, and the values have differing variances; the contextual information is often essential, and yet difficult to capture and incorporate in the classification process; many sources of noise, instrumental glitches, etc., can masquerade as transient events in the data stream; new, heterogeneous data arrive, and the classification must be iterated dynamically. Requiring a high completeness (don't miss any interesting events) and low contamination (a few false alarms), and the need to complete the classification process and make an optimal decision about expending

---

[1] California Institute of Technology, [george,donalek,aam]@astro.caltech.edu
[2] Jet Propulsion Laborattory, [baback,turmon]@jpl.nasa.gov
[3] California Institute of Technology, [mjg,ajd]@cacr.caltech.edu
[4] California Institute of Technology, [nihar,cheny]@caltech.edu
[5] Distinguished Visiting Professor, King Abdulaziz University, Jeddah, Saudi Arabia

valuable follow-up resources (e.g., obtain additional measurements using a more powerful instrument at a certain cost) in real time are challenges that require some novel approaches.

While this situation arises in many domains, it is especially true for the developing field of time domain astronomy. Telescope systems are dedicated to discovery of moving objects (e.g., potentially hazardous, Earth-crossing asteroids [1,2,3], transient or explosive astrophysical phenomena, e.g., supernovae (SNe), γ-ray bursts (GRBs), etc. – each requiring rapid alerts and follow-up observations. The time domain is rapidly becoming one of the most exciting new research frontiers in astronomy [23,29], broadening substantially our understanding of the physical universe, and perhaps lead to a discovery of previously unknown phenomena [16,23,24].

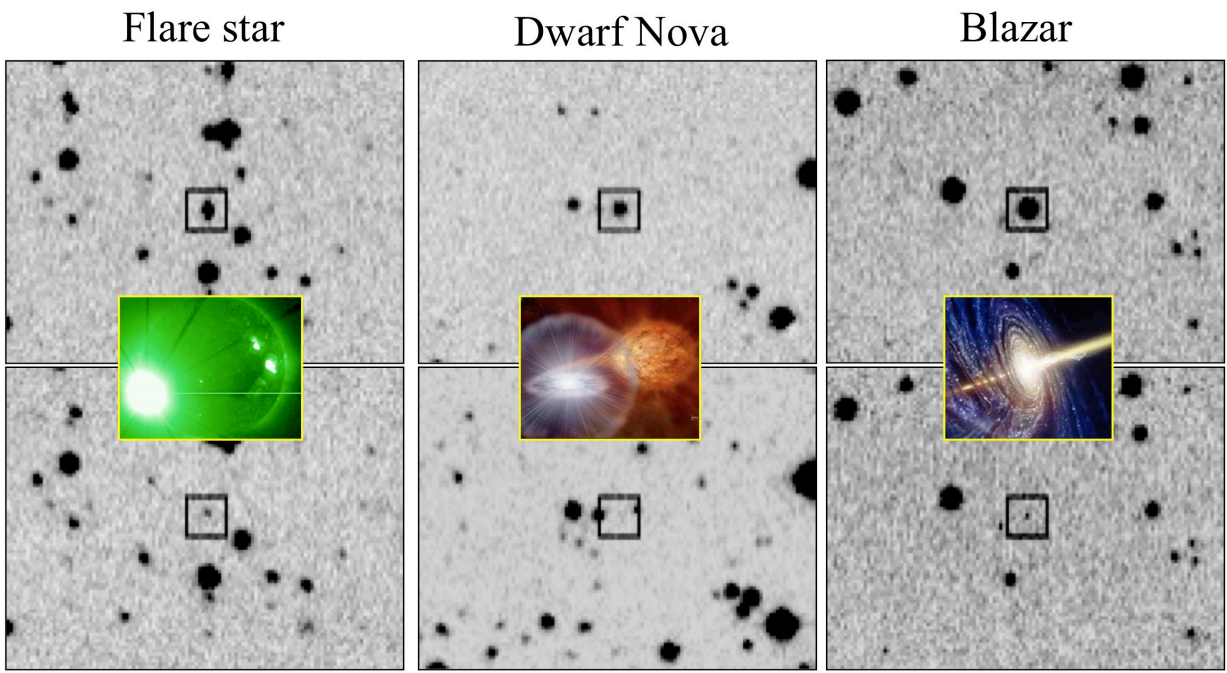

Figure 1. Examples of transient events from the Catalina Real-time Transient Survey (CRTS) sky survey [17,25]. Images in the top row show objects which appear much brighter that night, relative to the baseline images obtained earlier (bottom row). On this basis alone, the three transients are observationally indistinguishable, yet the subsequent follow-up shows them to be three vastly different types of phenomena: a flare star (left), a cataclysmic variable (dwarf nova) powered by an accretion to a compact stellar remnant (middle), and a blazar, flaring due to instabilities in a relativistic jet (right). Accurate transient event classification is the key to their follow-up and physical understanding.

The key to progress in time-domain astrophysics is the availability of substantial event data streams generated by panoramic digital synoptic sky surveys, coupled with a rapid follow-up of potentially interesting events (photometric, spectroscopic, and multi-wavelength). Physical classification of the transient sources is the key to their interpretation and scientific uses, and in many cases scientific returns come from the follow-up observations that depend on scarce or costly resources (e.g., observing time at larger telescopes). Since the transients change rapidly, a rapid (as close to the real time as possible) classification, prioritization, and follow-up are essential, the time scale depending on the nature of the source, which is initially unknown. In some cases the initial classification may remove the rapid-response requirement, but even an archival (i.e., not time-critical) classification of transients poses some interesting challenges.

A number of synoptic astronomical surveys are already operating [see, e.g., 1,2,3,7,17,25,26,43], and much more ambitious enterprises [4,5] will move us into the Petascale regime, with hundreds of thousands of transient events per night, implying a need for an automated, robust processing and follow-up, sometimes using robotic telescopes. Thus, *a new generation of scientific measurement systems is emerging* in astronomy, and many other fields: connected sensor networks which gather and analyze data automatically, and respond to outcome of these measurements in the real-time, often redirecting the measurement process itself, and without human intervention.

We are developing a novel set of techniques and methodology for an automated, real-time data analysis and discovery, operating on massive and heterogeneous data streams from robotic telescope sensor networks, fully integrated with Virtual Observatory (VO) [39,40,42]. The system incorporates machine learning elements for an iterative, dynamical classification of astronomical transient events, based on the initial detection measurements, archival information, and newly obtained follow-up measurements from robotic telescopes. A key novel feature, still under development, will be the ability to define and request particular types of follow-up observations in an automated fashion. Our goal is to increase the efficiency and productivity of a number of synoptic sky survey data streams, and enable new astrophysical discoveries.

## 2. THE CHALLENGE OF AN AUTOMATED, REAL-TIME EVENT CLASSIFICATION

A full scientific exploitation and understanding of astrophysical events requires a rapid, multi-wavelength follow-up. The *essential enabling technologies* that need to be automated are robust classification and decision making for the optimal use of follow-up facilities. They are the key for exploiting the full scientific potential of the ongoing and forthcoming synoptic sky surveys.

The first challenge is to associate classification probabilities that any given event belongs to a variety of known classes of variable astrophysical objects and to update such classifications as more data come in, until a scientifically justified convergence is reached [24]. Perhaps an even more interesting possibility is that a given transient represents a previously unknown class of objects or phenomena, that may register as having a low probability of belonging to any of the known data models. The process has to be *as automated as possible, robust, and reliable*; it has to operate from *sparse and heterogeneous data*; it has to maintain a *high completeness* (not miss any interesting events) yet a *low false alarm rate*; and it has to *learn* from the past experience for an ever improving, evolving performance. The next step is development and implementation of an automated follow-up event prioritization and decision making mechanism, which would actively determine and request follow-up observations on demand, driven by the event data analysis. This would include an automated identification of the most discriminating potential measurements from the available follow-up assets, taking into account their relative cost functions, in order to optimize both classification discrimination, and the potential scientific returns.

An illustration of an existing, working system for a real-time classification of astrophysical event candidates in a real synoptic sky survey context is shown in Fig. 2. This is an Artificial Neural Network (ANN) based classifier [18] that separates real transient sources from a variety of spurious candidates caused by various data artifacts (electronic glitches, saturation, cross-talk, reflections, etc.), that operated as a part of the Palomar-Quest (PQ) survey's [7,26] real time data reduction pipeline. While this is a very specialized instance of an automated event classifier for a particular sky survey experiment, it illustrates the plausibility and the potential of this concept. A similar approach, using Support Vector Machine (SVM) techniques [11], has been deployed successfully by the Lawrence Berkeley National Laboratory Nearby Supernova Factory [10,27]. Use of image morphology for astronomical image classification via machine learning has long been used successfully, e.g., [12,19,20]. Here we deploy it in a real-time data reduction pipeline.

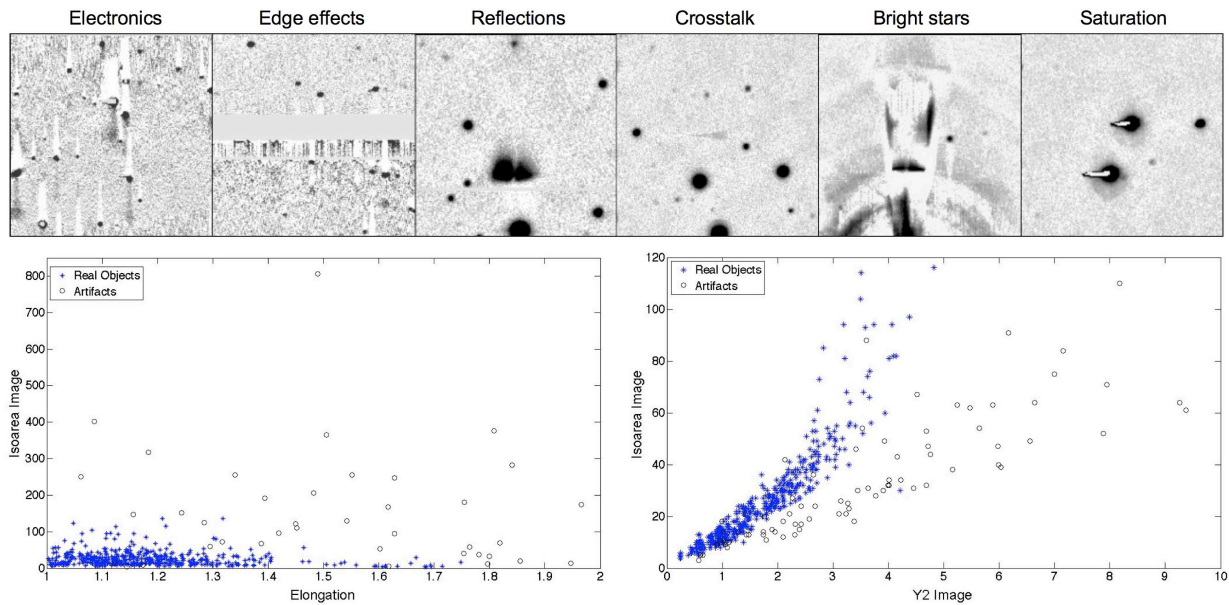

Figure 2. Automated classification of candidate events, separating real astronomical sources from a variety of spurious candidates (instrument artifacts) is operational within the Palomar-Quest) survey's real time data pipeline [26,31]. Image cutouts on the top show a variety of instrumental and data artifacts which appear as spurious transients, since they are not present in the baseline comparison images. The two panels on the bottom show a couple of morphological parameter space projections, in which artifacts (O) separate well from genuine objects (✷). A multi-layer perceptron (MLP) ANN is trained to separate them, using 4 image parameters, with an average accuracy of ~ 95%. From Donalek et al., [31] and in prep.

However, the problem here is more complex and challenging: it is an astrophysical classification of genuine transient events, all of which would look the same in the images (star-like), so that information other than image morphology must be used.  One problem is that in general, not all parameters would be measured for all events, e.g., some may be missing a measurement in a particular filter, due to a detector problem; some may be in the area on the sky where there are no useful radio observations; etc.  Broader approaches to automated classification of transients and variables include, e.g., [28,30,31,44,45,46,47,48].

A more insidious problem is that many observables would be given as upper or lower limits, rather than as well defined measurements; for example, "the increase in brightness is > 3.6 magnitudes", or "the radio to optical flux ratio of this source is < 0.01".  One approach is to treat them as missing data, implying a loss of the potentially useful information.  A better approach is to reason about "censored" observations, that can be naturally incorporated through a Bayesian model by choosing a likelihood function that rules out values violating the bounds.

## 3. A BAYESIAN APPROACH TO EVENT CLASSIFICATION

We identify two core problems: *classification* (physical interpretation of an event), learning from compiled knowledge obtained by linking observations to phenomena, and *recommendation* (what are the optimal follow-up observations for this particular event).

The main astronomical inputs are in the form of observational and archival parameters for individual objects, which can be put into various, often independent subsets. Examples include fluxes measured at different wavelengths, associated colors or hardness ratios, proximity values, shape measurements, magnitude characterizations at different timescales, etc. The heterogeneity and sparsity of data makes the use of Bayesian methods for classification a natural choice.

Distributions of such parameters need to be estimated for each type of variable astrophysical phenomena that we want to classify.  Then an estimated probability of a new event belonging to any given class can be evaluated from all of such pieces of information available, as follows. Let us denote the feature vector of event parameters as $x$, and the object class that gave rise to this vector as $y$, $1 \leq y \leq K$.  While certain fields within $x$ will generally be known, such as sky position and brightness in selected filters, many other parameters will be known only sporadically, e.g., brightness change over various time baselines.  In a Bayesian approach, $x$ and $y$ are related via

$$P(y = k \mid x) = P(x \mid y = k)P(k)/P(x) \propto P(k)P(x \mid y = k) \approx P(k)\prod_{b=1}^{B} P(x_b \mid y = k)$$

Because we are only interested in the above quantity as a function of $k$, we can drop factors that only depend on $x$.  We assume that, conditional on the class $y$, the feature vector *decomposes* into $B$ roughly independent blocks, generically labeled $x_b$.  These blocks may be singleton variables, or contain multiple variables, e.g., sets of filters that are highly correlated.  The resulting algorithm is called *naive Bayes* because of its assumption that we may decouple the inputs in this way [8,9].

This decoupling is advantageous to us in two ways. First, it allows us to circumvent the "curse of dimensionality," because we will eventually have to learn the conditional distributions $P(x_b \mid y = k)$ for each $k$. As more components are added to $x_b$, more examples will be needed to learn the corresponding distribution. The decomposition keeps the dimensionality of each feature block manageable. Second, such decomposition allows us to cope easily with ignorance of missing variables. We simply drop the corresponding factors from the product above.

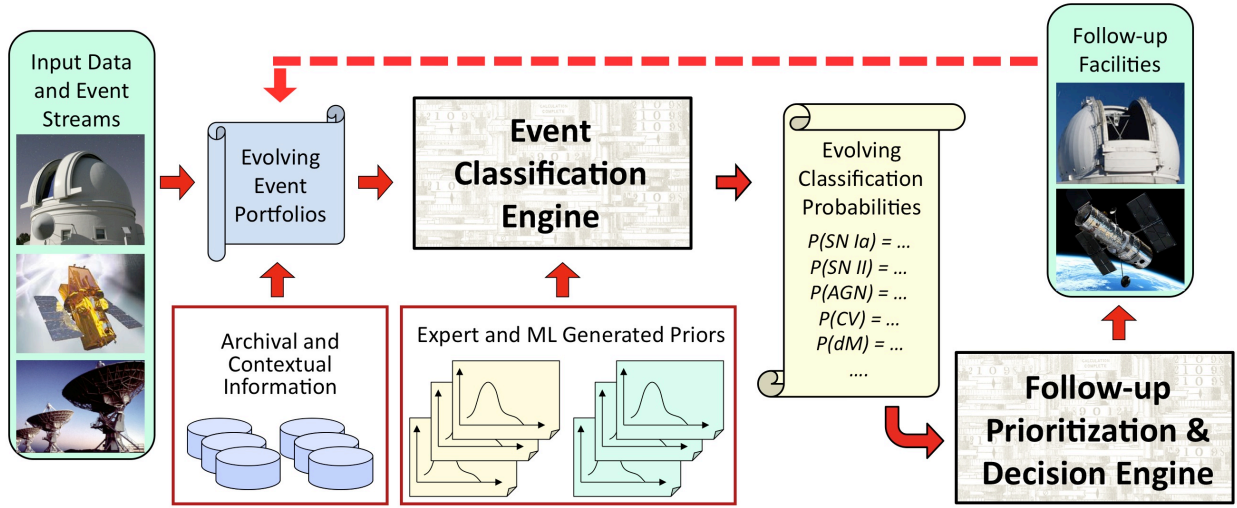

Figure 3. A conceptual outline of the system. The initial input consists of the generally sparse data describing transient events discovered in sky surveys, supplemented by archival heterogeneous measurements from external, multi-wavelength archives corresponding to this spatial location, if available (e.g. radio flux and distance to nearest galaxy). Data are collected in evolving electronic portfolios containing all currently available information for a given event. These data are fed into the Event Classification Engine; another input into the classification process is an evolving library of priors giving probabilities for observing these particular parameters if the event was belonging to a class X. The output of the classification engine is an evolving set of probabilities of the given event belonging to various classes of interest, which are updated as more data come in, and classifications change. This forms an input into the Follow-up Prioritization and Decision Engine, which would prioritize the most valuable follow-up measurements given a set of available follow-up assets (e.g., time on large telescopes, etc.), and their relative cost functions. What is being optimized is: (a) the new measurements which would have a maximum discrimination for ambiguous classifications, and/or (b) the follow-up measurements which would likely yield most interesting science, given the current best-guess event classification? New measurements from such follow-up observations will be fed back into the event portfolios, leading to dynamically updated/iterated classifications, repeating the cycle.

As a simple demonstration of the technique, we have been experimenting with a prototype Bayesian Network (BN) model [32,33]. We use a small but homogeneous data set involving colors of ~ 1,000 reliably classified transients detected in the CRTS survey [17,25], as measured at the Palomar 1.5-m telescope. We have used multinomial nodes (discrete bins) for 3 colors, with provision for missing values, and a multinomial node for Galactic latitude which is always present and is a probabilistic indicator of whether an object is Galactic or not. The current priors used are for six distinct classes, cataclysmic variables (CVs; these are binary star system in which a compact stellar remnant such as a white dwarf or a neutron star accretes material from its companion in a fairly stochastic fashion), supernovae (SN; these are exploding stars, and while there are several distinct types, the overall behavior is very similar), blazars (beamed active galactic nuclei, or AGN, where we are looking into their relativistic jet), other variable AGN, UV Ceti stars (dwarf stars undergoing gigantic equivalent of the Solar flares), and all else bundled into a sixth pseudo-class, called Rest. Testing is done with a 10-fold cross validation, in order to assess how good it will perform on an independent data set.

Using a sample of 316 SNe, 277 CVs, and 104 blazars, and a *single* epoch measurement of colors, in the relative classification of CVs vs. SNe, we obtain a completeness of ~ 80% and a contamination of ~ 19%, which reflects a qualitative color difference between these two types of transients. In the relative classification of CVs vs. blazars, we obtain a completeness of ~ 70 – 90% and a contamination of ~ 10 – 24% (the ranges corresponding to different BN experiments), which reflects the fact that colors of these two types of transients tend to be similar, and that some additional discriminative parameter is needed. Eventually we will use a BN with an order of magnitude more classes, including divisions of different types of SNe, AGN, and a large variety of variable star types (there are literally hundreds of varieties of variable stars, but only a few tens may be relevant for the present transients search), with more measured parameters, and additional BN layers. Measurements from multiple epochs should improve considerably the classifications. The end result will be the posteriors for the "Class" node from the marginalized probabilities of all available inputs for a given object.

In this framework the priors come from a set of observed parameters like distribution of colors, distribution of objects as a function of Galactic latitude, frequencies of different types of objects etc. The posteriors we are interested in are determining the type of an object based on, say, its (*r-i)* color, Galactic latitude and proximity to another object etc.

Sparse and/or irregular light curves (LC) from any given object class can have sufficient salient structure that can be exploited by automated classification algorithms. We have experimented with Gaussian Process Regression [34], and found it to be useful for parameter estimation for a certain types of LCs that can be represented by a standard data model (e.g., Supernovae).

We are now experimenting with a different approach. By *pooling* many instances of an object class's LCs we can effectively represent and encode their characteristic structure *probabilistically*, and construct an empirical probability distribution function (PDF) that can be used for subsequent classification of new event observations. This comparison can be made incrementally over time as new observations "trickle in", with the final classification scores growing more confident with each additional set of observations that is accumulated.

Since the telescope's (flux-only) observations come primarily in the form of single magnitude changes over time increments – *e.g.*, an observed (Δt, Δm) pair – we focus on modeling the joint distribution of all such pairs of data points for a given LC (Note: we consider all possible *causal* increments available, corresponding to Δt > 0). By virtue of being *increments*, these data and their empirical PDF will be invariant to absolute magnitude (the distance to the event generally being unknown) and time (the onset of the event not being known) shifts. Additionally, these densities allow flux upper limits to be encoded as well – *e.g.*, under poor seeing conditions, we may only obtain bounded observations such as m > 18. We currently use smoothed 2D histograms to model the distribution of (Δt, Δm) pairs. This is a computationally simple, yet effective way to implement a non-parametric density model that is flexible enough for all object classes under our consideration. Figure 4 shows the joint 2D histograms for 3 classes of objects and how a given probe LC measurements fit these 3 class-specific histograms.

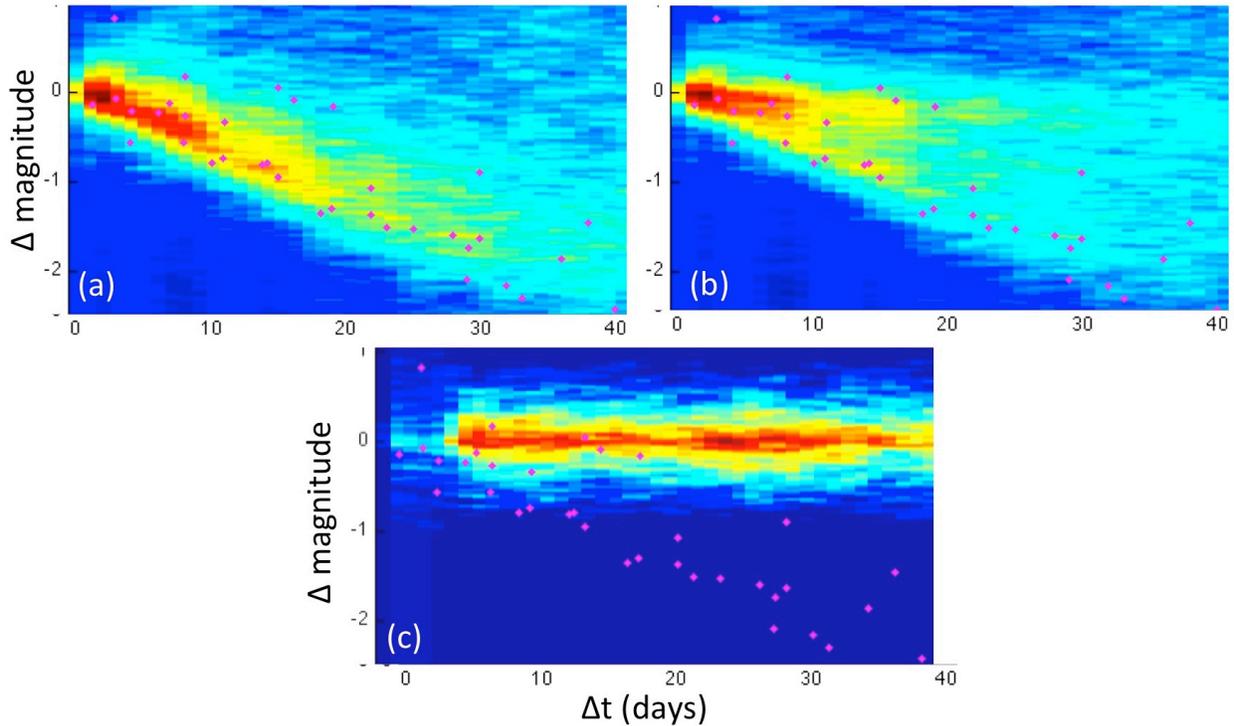

Figure 4. Examples of (Δm, Δt) pairs PDFs for three types of astrophysical transients: (a) SN Ia, (b) SN IIP and (c) RR Lyrae, using bins of width Δt = 1 day, and Δm = 0.01. The histograms were smoothed with a 3-tap triangular Δt kernel = [0.25 0.5 0.25] and a Gaussian Δm kernel of FWHM = 0.05 mag. The set of diamonds superimposed on each panel are from a single test case of a SN Ia's LC. Note that PDFs for the two SN types form a better "fit" to the observed data (diamonds) than the RR Lyrae's PDF (and SN Ia is a better fit than SN II P). Various metrics on probability distributions can be used to automatically quantify the degree of fitness.

In our preliminary experimental evaluations with a small number of object classes (single outburst like SN, periodic variable stars like RR Lyrae and Miras, as well as stochastic like blazars and CVs) we have been able to show that our gap event density models are potentially a powerful classification method from sparse/irregular time series like typical observational LC data.

## 4. INCORPORATING THE CONTEXTUAL INFORMATION

Contextual information can be highly relevant to resolving competing interpretations: for example, the light curve and observed properties of a transient might be consistent with both it being a cataclysmic variable star, a blazar, or a supernova. If it is subsequently known that there is a galaxy in close proximity, the supernova interpretation becomes much more plausible. Such information, however, can be characterized by high uncertainty and absence, and by a rich structure – if there were two candidate host galaxies, their morphologies, distance, etc., become important, e.g., is this type of supernova more consistent with being in the extended halo of a large spiral galaxy or in close proximity to a faint dwarf galaxy? The ability to incorporate such contextual information in a quantifiable fashion is highly desirable. In a separate project we are investigating the use of crowdsourcing as a means of harvesting the human pattern recognition skills, especially in the context of capturing the relevant contextual information, and turning them into machine-processable algorithms.

A methodology employing contextual knowledge forms a natural extension to the logistic regression and classification methods mentioned above. Ideally such knowledge can be expressed in a manipulable fashion within a sound logical model, for example, it should be possible to state the rule that "a supernova has a stellar progenitor and will be substantially brighter than it by several order of magnitude" with some metric of certainty and infer the probabilities of observed data matching it. *Markov Logic Networks* (MLNs, [36]) are such a probabilistic framework using declarative statements (in the form of logical formulae) as atoms associated with real-valued weights expressing their strength. The higher the weight, the greater the difference in log probability between a world that satisfies the formula and one that does not, all other thing being equal. In this way, it becomes possible to specify 'soft' rules that are likely to hold in the domain, but subject to exceptions - contextual relationships that are likely to hold such as supernovae may be associated with a nearby galaxy or objects closer to the Galactic plane may be stars.

A MLN defines a probability distribution over possible worlds with weights that can be learned generatively or discriminatively: it is a model for the conditional distribution of the set of query atoms *Y* given the set of evidence atoms *X*. Inferencing consists of finding the most probable state of the world given some evidence or computing the probability that a formula holds given a MLN and set of constants, and possibly other formulae as evidence. Thus the likelihood of a transient being a supernova, depending on whether there was a nearby galaxy, can be determined.

The structure of a MLN – the set of formulae with their respective weights – is also not static but can be revised or extended with new formulae either learned from data or provided by third parties. In this way, new information can easily be incorporated. Continuous quantities, which form much of astronomical measurements, can also be easily handled with a hybrid MLN [37].

## 5. COMBINING AND UPDATING THE CLASSIFIERS

An essential task is to derive an optimal event classification, given inputs from a diverse set of classifiers such as those described above. This will be accomplished by a fusion module, currently under development, illustrated schematically in Fig. 5.

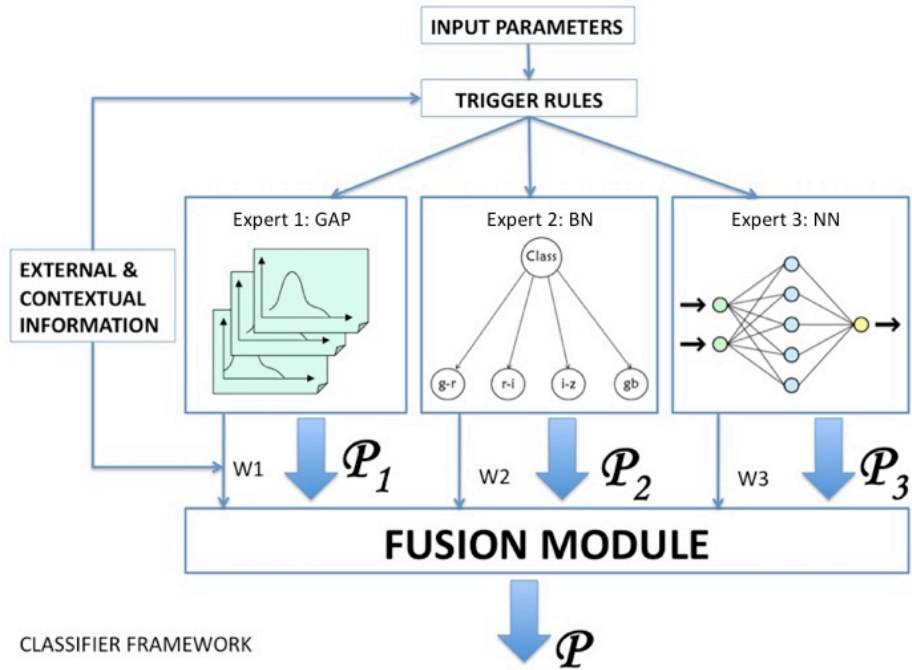

Figure 5. A schematic illustration of the event classifier combination challenge, to be implemented by the classification fusion module. Different aspects of available event information trigger different classifiers. In some cases more than one classifier can be used. How to combine the different outcomes is a subject of the ongoing work.

A MLN approach could be used to represent a set of different classifiers and the inferred most probable state of the world from the MLN would then give the optimal classification. For example, a MLN could fuse the beliefs of different ML-based transient classifiers – 4 give a supernova and 3 give a cataclysmic variable, say – to give a definitive answer.

We are experimenting with the so-called "sleeping expert" [35] method. A set of different classifiers each generally works best with certain kinds of inputs. Activating these optionally only when those inputs are present provides an optimal solution to the fusion of these classifiers. Sleeping expert can be seen as a generalization of the IF-THEN rule: IF this condition is satisfied THEN activate this expert, e.g., a specialist that makes a prediction only when the instance to be predicted falls within their area of expertise. For example, some classifiers work better when certain inputs are present, and some work only when certain inputs are present. It has been shown that this is a powerful way to decompose a complex classification problem. External or *a priori* knowledge can be used to awake or put experts to sleep and to modify online the weights associated to a given classifier; this contextual information may be also expressed in text.

A crucial feature of the system should be the ability to update and revise the prior distributions on the basis of the actual performance, as we accumulate the true physical classifications of events, e.g., on the basis of follow-up spectroscopy. Learning, in the Bayesian view, is precisely the action of determining the probability models above – once determined, the overall model (1) can be used to answer many relevant questions about the events. Analytically, we formulate this as determining unknown distributional parameters θ in parameterized versions of the conditional distributions above, $P(x \mid y = k; \theta)$. (Of course, the parameters depend on the object class *k*, but we suppress this below.) In a histogram representation, θ is just the probabilities associated with each bin, which may be determined by computing the histogram itself. In a Gaussian representation, θ would be the mean vector μ and covariance matrix Σ of a multivariate Gaussian distribution, and the parameter estimates are just the corresponding mean and covariance of the object-*k* data. When enough data is available we can adopt a semi-parametric representation in which the distribution is a linear superposition of such Gaussian distributions,

$$P(x_d \mid y = k; \theta) = \sum_{m=1}^{M} \lambda_m N(x_d; \mu_m, \Sigma_m)$$

This generalizes the Gaussian representation, since by increasing *M*, more distributional characteristics may be accounted for. The corresponding parameters may be chosen by the Expectation-Maximization algorithm [13]. Alternatively, kernel density estimation could be used, with density values compiled into a lookup table [14,21].

We can identify three possible sources of information that can be used to find the unknown parameters. They can be from the *a priori* knowledge, e.g. from physics or monotonicity considerations, or from examples that are labeled by experts, or from the feedback from the downstream observatories once labels are determined. The first case would serve to give an analytical form for the distribution, but the second two amount to the provision of labeled examples, (*x*, *y*), which can be used to select a set of *k* probability distributions.

## 6. AUTOMATED DECISION MAKING FOR AN OPTIMIZED FOLLOW-UP

We typically have sparse observations of a given object of interest, leading to classification ambiguities among several possible object types (e.g., when an event is roughly equally likely to belong to two or more possible object classes, or when the initial data are simply inadequate to generate a meaningful classification at all). Generally speaking, some of them would be of a greater scientific interest than others, and thus their follow-up observations would have a higher scientific return. Observational resources are scarce, and always have some cost function associated with them, so a key challenge is to determine the follow-up observations that are most useful for improving classification accuracy, and detect objects of scientific interest.

There are two parts to this challenge. First, what type of a follow-up measurement – given the *available* set of resources (e.g., only some telescopes/instruments may be available) – would yield the maximum information gain in a particular situation? And second, if the resources are finite and have a cost function associated with them (e.g., you can use only so many hours of the telescope time), when is the potential for an interesting discovery worth spending the resources?

We take an information-theoretic approach to this problem [15] that uses Shannon entropy to measure ambiguity in the current classification. We can compute the entropy drop offered by the available follow-up measurements – for example, the system may decide that obtaining an optical light curve with a particular temporal cadence would discriminate between a Supernova and a flaring blazar, or that a particular color measurement would discriminate between, say, a cataclysmic variable eruption and a gravitational microlensing event. A suitable prioritized request for the best follow-up observations would be sent to the appropriate robotic (or even human-operated) telescopes.

Note that the system is suggesting follow-up observations that may involve imperfect observations of a block of individual variables. This is a more powerful capability than rank-ordering individual variables regarding their helpfulness. Furthermore, we will ascertain that the framework accounts for the varying degrees of accuracy of different observations. The key to quantifying the classification uncertainty is the conditional entropy of the posterior distribution for $y$, given all the available data. Let $H[p]$ denote the Shannon entropy of the distribution $p$, which is always a distribution over object-class $y$. (The classification is discrete, so we only need to compute entropies of discrete distributions.) Then, when we take an additional observation $x_+$, uncertainty drops from $H[p(y \mid x_o)]$ to $H[p(y \mid x_o, x_+)]$. We want to choose the source $x_+$ so that the expected final entropy is lowest. To choose the best refinement in advance, we look for the largest expected drop in entropy.

Because all observing scenarios start out at the same entropy $H[p(y \mid x_o)]$, maximizing entropy drop is the same as minimizing expected final entropy, $E[H[p(y \mid x_o, x_+)]]$. The expectation is with respect to the distribution of the new variable $x_+$, whose value is not yet known. Therefore, this entropy is a function of the *distribution* of $x_+$, but not the value of the random variable $x_+$. The distribution captures any imprecision and noise in the new observation. In our notation, the best follow-on observation thus minimizes, over available variables $x_+$,

$$H[p(y \mid x_+, x_0)] = -\sum_{y, x_+} p(y, x_+ \mid x_0) \log p(y \mid x_+, x_0).$$

This is equivalent to maximizing the conditional mutual information of $x_+$ about $y$, given $x_o$; that is, $I(y; x_+ \mid x_o)$ [22]. *The density above is known within the context of our assumed statistical model.* Thus, we can compute, within the context of the previously learned statistical model, a rank-ordered list of follow-on observations, which will lead to the most efficient use of resources.

Alternatively, instead of maximizing the classification accuracy, we consider a scenario where the algorithm chooses a set of events for follow-up and subsequent display to an astronomer. The astronomer then provides information on how interesting the observation is. The goal of the algorithm is to learn to choose follow-up observations which are considered most interesting. This problem can be naturally modeled using *Multi-Armed Bandit* algorithms (MABs) [38]. The MAB problem can abstractly be described as a slot machine with *k* levers, each of which has different expected returns (unknown to the decision maker). The aim is to determine the best strategy to maximize returns. There are two extreme approaches: (1) exploitation - keep pulling the lever which, as per your current knowledge, returns most, and (2) exploration – experiment with different levers in order to gather information about the expected returns associated with each lever. They key challenge is to trade off exploration and exploitation. There are algorithms [47] guaranteed to determine the best choice as the number of available tries goes to infinity.

In this analogy different telescopes and instruments are the levers that can be pulled. Their ability to discriminate between object classes forms the returns. This works best when the priors are well assembled and a lot is already known about the type of object one is dealing with. But due to the heterogeneity of objects, and increasing depth leading to transients being detected at fainter levels, and more examples of relatively rarer subclasses coming to light, treating the follow-up telescopes as a MAB will provide a useful way to rapidly improve the classification and gather more diverse priors. An analogy could be that of a genetic algorithm which does not get stuck in a local maxima because of its ability to sample a larger part of the parameter space.

This work is supported in part by the NASA grant 08-AISR08-0085, the NSF grant AST-0909182, and the U.S. Virtual Astronomical Observatory; NS and YC were supported in part by the Caltech SURF program. We thank numerous collaborators and colleagues for stimulating discussions.